ub
\documentclass[a4paper,11pt]{article}
\pdfoutput=1 

\usepackage{jheppub} 

\usepackage[T1]{fontenc} 

\usepackage[latin1]{inputenc}
\usepackage{bm}
\usepackage{epsfig}
\usepackage{amsthm}
\usepackage{amsfonts}
\usepackage{float}
\usepackage{amsmath,amssymb}
\usepackage{color}
\usepackage{slashed}
\usepackage{relsize}
\usepackage{bbold}
\usepackage{graphicx}
\usepackage{dcolumn}
\usepackage{bm}
\usepackage[normalem]{ulem}
\usepackage{cancel}
\usepackage{tikz}

\newcommand{\nn}{\nonumber}
\newcommand{\be}{\begin{equation}}
\newcommand{\ee}{\end{equation}}
\newcommand{\bea}{\begin{eqnarray}}
\newcommand{\eea}{\end{eqnarray}}

\title{\boldmath Thermal Field Theory of the Tsallis statistics}

\author[a,b]{Mahfuzur Rahaman,}
\author[c]{Trambak Bhattacharyya,}
\author[a,b]{Jan-e Alam}

\affiliation[a]{Variable Energy Cyclotron Centre, 
1/AF, Bidhan Nagar, Kolkata - 700064, India}
\affiliation[b]{Homi Bhabha National Institute, Anushaktinagar, Mumbai, India}
\affiliation[c]{Bogoliubov Laboratory of Theoretical Physics, Joint Institute for Nuclear Reseach, Dubna, Russia}

\emailAdd{mahfuzurrahaman01@gmail.com}
\emailAdd{bhattacharyya@theor.jinr.ru}
\emailAdd{jane@vecc.gov.in}

\abstract{Classical and quantum Tsallis distributions have been widely used in many branches of natural and social sciences. 
But, the quantum field theory of the Tsallis distributions is relatively a less explored arena. In this article, we 
derive the expression for the thermal two-point functions for the Tsallis statistics with the help of the corresponding 
statistical mechanical formulations. We show that the quantum Tsallis distributions used in the 
literature appear in the thermal part of the propagator much in the same way the Boltzmann-Gibbs distributions appear in 
the conventional thermal field theory. As an application of our findings, we calculate the thermal mass in the $\phi^4$ scalar field theory 
within the realm of the Tsallis statistics. }

\begin{document} 
\maketitle
\flushbottom

\section{Introduction}
\label{sec:intro}
In spite of a remarkable success of the Boltzmann-Gibbs (BG) statistical mechanics, 
there exist many systems which may not be describable by this formulation  \cite{tsallisbook,tsallisgellmann}. Some
generalized concepts are needed to deal with such systems. These systems often contain fluctuations 
(of temperature, number density etc.) inside their boundary, and/or they may experience long-range correlation.
It has been established \cite{wilk,wilkprc,wilkchaos,biroprl} that these kinds of systems may be understood from a generalization of the 
Boltzmann-Gibbs formulations developed by C. Tsallis \cite{tsallis88}, and is known as the `Tsallis statistics'.

Since then, the Tsallis-like single particle distributions have been used in many branches of natural \cite{plastinopla,olbert,tsalliscosmic,beckcosmic,rafelskidiffu,tbsmdiffu,chem,compu,bio,bhupalastro} and social sciences \cite{eco,cogni,lingu}, 
and the field of high energy collisions is no exception
\cite{PHENIX1,CMS1,CMS2,ALICE_deuteron,Biro09,Cleymans09,Cleymans12,Cleymans13,khandai,TsallisTaylor,tsallisraa,Grigoryan17,lacey,ishihara,bcmmp,bcmpst,Azmi14,deppmanYMTsFract1,deppmanYMTsFract2}. 
These studies have now established that the 
a global observable like the hadronic transverse momentum distribution generated from the high energy 
collisions (of protons on protons, for example) is describable by the `Tsallis-like' distributions up to
a very high range of transverse momentum ($p_{\text{t}}$) \cite{Azmi14}. Of particular importance is the power-law 
phenomenological distribution function below \cite{Cleymans12,bcmmp},
\begin{equation}
\frac{d^2N}{p_{\text{t}}dp_{\text{t}}dy}= \mathcal{C} \omega \big[1+\beta(q-1)\omega \big]^{-\frac{q}{q-1}},
\label{tsallisMBn0}
\end{equation}
where $\mathcal{C}$ is a constant factor, $\omega$ is the single particle energy, and $y$ is rapidity.
Many of the above-mentioned articles utilize this distribution to describe the experimental data. 
The distribution in eq.~\eqref{tsallisMBn0}
is popularly known as the `Tsallis distribution' which is described by the entropic parameter $q$ 
and the Tsallis 
(inverse) temperature $\beta$. It has been established in ref.~\cite{wilk} that in a system of fluctuating temperature 
zones, $\beta$ is the average inverse temperature and $q$ is related to the relative variance in temperature. 
In the limit $q\rightarrow 1$, eq.~\eqref{tsallisMBn0} becomes Boltzmann like,
\begin{equation}
\frac{d^2N}{p_{\text{t}}dp_{\text{t}}dy}= \mathcal{C} \omega e^{-\beta \omega},
\end{equation}
and according to ref.~\cite{wilk}, this implies a gradual disappearance of fluctuations inside the system.

There have been attempts to verify whether the phenomenological Tsallis distribution belongs to the exact Tsallis 
statistical mechanical formulations. Recently, it has been shown in \cite{Parvan19} that the transverse momentum spectra 
obtained from the exact Tsallis statistical mechanics is expressible in terms of a series expansion, 
and the zeroth order approximation of that series (for Maxwell-Boltzmann particles) is the distribution given by eq.~\eqref{tsallisMBn0}. 
%

In the present article, we are, however, interested in the quantum extensions of the Tsallis distribution. 
The forms of the quantum distributions we concentrate upon are given below \cite{Hasegawa},
\begin{eqnarray}
&\text{Tsallis~Fermi-Dirac:~}& 
\label{TFD}
\frac{1}{\big[1+\beta(q-1)\omega\big]^{\frac{q}{q-1}}+1} \\
&\text{Tsallis~Bose-Einstein:~}& 
\frac{1}{\big[1+\beta(q-1)\omega\big]^{\frac{q}{q-1}}-1}.
\label{TBE}
\end{eqnarray}
Many other studies \cite{conroypla,buyupla,penpla,millerpla,millerprd,Mitra18} 
use a slightly different form of the Tsallis quantum distributions given by,
\begin{eqnarray}
\frac{1}{\big[1+\beta(q-1)\omega\big]^{\frac{1}{q-1}}\pm1},
\label{tsallisquaph}
\end{eqnarray}
where positive(negative) sign appears for the fermions(bosons). While eqs.~\eqref{TFD}, and~\eqref{TBE} can be obtained 
from the statistical mechanical formulations proposed by Tsallis \cite{Tsallis3}, eq.~\eqref{tsallisquaph} fails to show this connection. 
We will show that as a  quantum extension of eq.~\eqref{tsallisMBn0} 
our results are consistent with the expressions given in eqs.~\eqref{TFD} and \eqref{TBE}. 
It is worth observing that the Tsallis quantum distribution approaches the corresponding Boltzmann-Gibbs 
quantum distribution when $q$ tends to 1. Also, in the high temperature regime ($q\neq 1$), 
both of them approach the classical limit given by eq.~\eqref{tsallisMBn0}.

\textcolor {black} It is well-known that the Boltzmann-Gibbs quantum distributions  appear
naturally through the two point functions evaluated using 
the techniques of the thermal quantum field theory 
(see e.g. \cite{bellac,das,kapusta,smallik}). In the present work, we calculate
the thermal two-point functions from the Tsallis generalization of thermal field theory in which the Tsallis
quantum distributions given by eqs.~\eqref{TFD}, and~\eqref{TBE} appear. The generalization of thermal field
theory in the Tsallis statistics has been addressed in ref.~\cite{niegawa}, which, however, adapts the Tsallis 
statistical mechanical formulation \cite{Tsallis3} that uses a definition of the expectation values different 
from that being used in the present paper 
(and in the phenomenological studies). Authors in ref.~\cite{olemskoi} also study the evolution of the most probable 
values of the order parameters by using the field theory of non-additive system.
However, the mentioned articles do not attempt to provide an explicit derivation of the quantum 
distributions given in eqs.~\eqref{TFD}, and~\eqref{TBE} which are of importance in the theoretical, 
phenomenological, and experimental studies. This is why in the present work we derive the quantum 
Tsallis distributions given by eqs.~\eqref{TFD}, and~\eqref{TBE} using the generalized techniques of thermal field theory. 
It is expected that the present work will address the need of a quantum field theory of the Tsallis 
distributions used in various branches of physics and beyond. 

%

The paper is organized as follows. In the next section the basic formulations of the Tsallis
statistical mechanics are discussed. In sections~\ref{sec:propboson}, and~\ref{sec:propfermion} the real time thermal 
propagators for scalar and  Dirac particles are derived respectively. The results obtained in section~\ref{sec:propboson}  
have been applied to estimate the thermal mass of real scalar bosons in section~\ref{sec:tsallismD}. Section~\ref{sec:summaconclu} 
is devoted to summary, discussions, and conclusions.


\section{Free Tsallis thermal propagator for the real scalar field theory}
\label{sec:propboson}

\begin{figure}[tbp]
\centering 
\hspace*{2cm} \includegraphics[width=.65\textwidth,trim=0 250 0 200,clip]{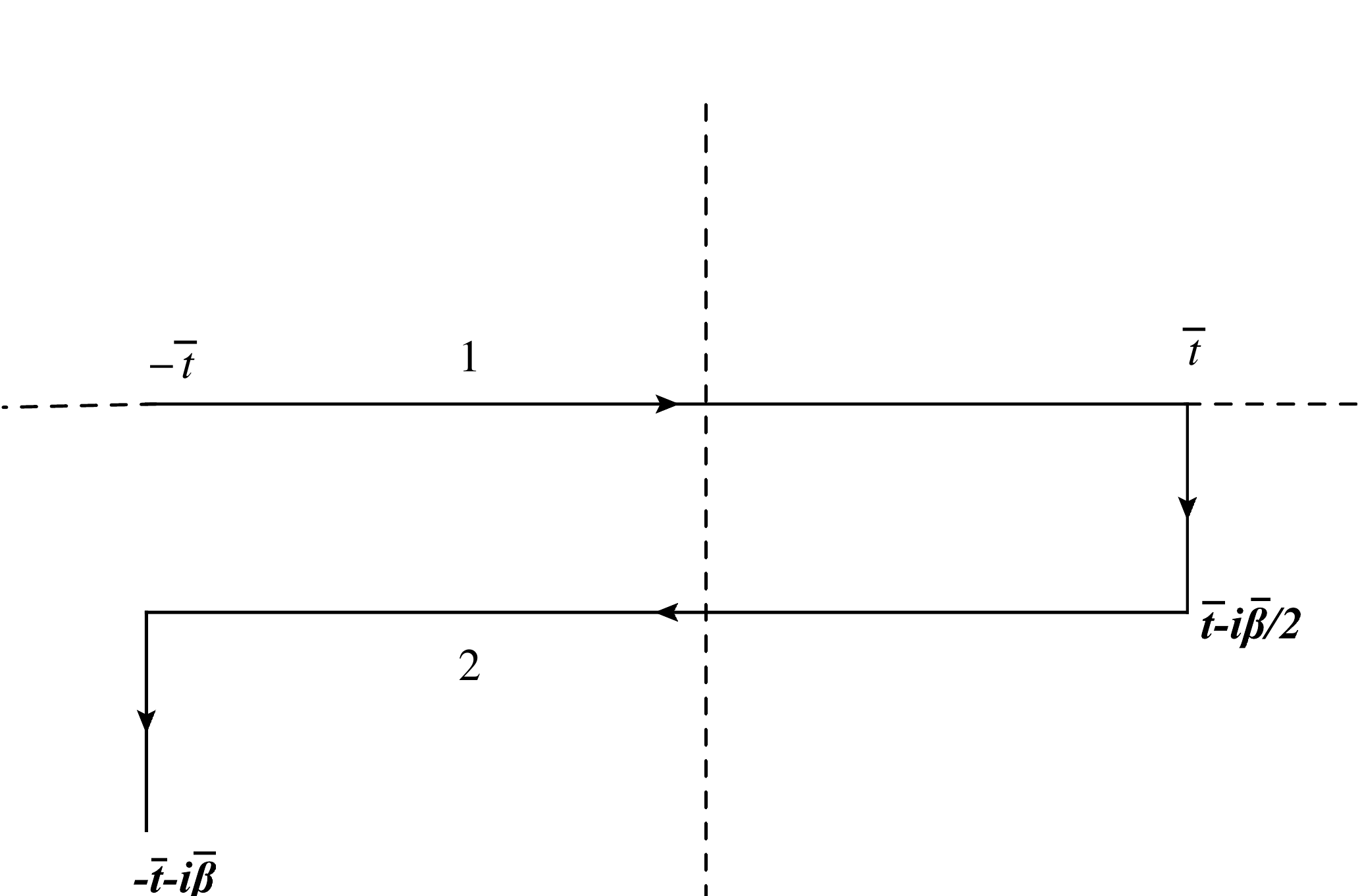}
\hfill
\caption{\label{fig1} Contour C on the time plane in the real time formalism. $\bar{\beta}$ is given by eq.~\eqref{betabar}.}
\end{figure}

\subsection{Some basic formulae and relations}
\label{subsec:formulae}
The Tsallis statistics begins with a generalized definition of entropy \cite{Tsallis3} which, 
if written in terms of density matrix $\hat{\rho}$, takes the following form,
\be
S_q\left[\hat{\rho}^q\right] = \frac{1-\text{Tr} \hat{\rho}^q}{q-1},
\ee
subject to the normalization condition,
\be
\text{Tr} \hat{\rho} =1.
\ee
The thermal expectation value is given by \cite{Tsallis3,Abe},
\begin{equation}
\langle\hat{A}\rangle_q=\text{Tr}\big( \left\{\hat{\rho}\right\}^q \hat{A}\big).
\end{equation}
$S_q$ approaches the Boltzmann-Gibbs-Shannon entropy given by $S = -\text{Tr} \hat{\rho} \ln \hat{\rho}$ when $q$ approaches 1.

By extrimizing the potential function (for chemical potential, $\mu=0$),
\be
\Omega=\langle \hat{H} \rangle - TS_q,
\ee
for temperature $T=\beta^{-1}$ with respect to density matrix, we get,
\be
\hat{\rho}=Z_q^{-1}\left[1+(q-1)\beta \hat{H}\right]^{\frac{1}{1-q}}.
\ee 
\\
\\
$\hat{H}$ is the Hamiltonian such that $\langle\hat{H}\rangle=E$ where $E$ is the energy of the system. Partition function $Z_q$
is given by,
\be
Z_q=\text{Tr}'\left[1+(q-1)\beta \hat{H}\right]^{\frac{1}{1-q}}.
\ee
The prime ($'$) in trace implies the Tsallis cut-off condition, which means that the trace is taken over the energy eigen values
which satisfy the criterion $\big[1+\beta(q-1)E\big]\ge0$. The Boltzmann-Gibbs density matrix $\hat{\rho}\propto 
e^{-\beta\hat{H}}$ can be obtained when $q\rightarrow 1$. Given the Tsallis cut-off condition, we can write the density operator $\hat{\rho}$ in terms of the $q$-exponential operator defined as follows,
\be
e_q^{-\beta \hat{H}} = \big[1+\beta(q-1)\hat{H}\big]^{\frac{1}{1-q}}.
\label{qexp}
\ee
The above equation implies,
\be
\hat{\rho}=Z_q^{-1}e_q^{-\beta \hat{H}}.
\ee
In the limit $q\rightarrow 1$, the $q$-exponential operator approaches the conventional exponential counterpart.


With the help of the generalized definition of the expectation values, the Tsallis thermal propagator for the real scalar field $\phi(x)$ (with $x=(\tau,\vec{x})$) on a contour C can be defined as, 

\begin{eqnarray}
D(x,x') &=& i \langle \hat{T}_c \phi(x )\phi( x') \rangle_q \nn\\
&=& i\theta_c(\tau -\tau') \langle  \phi(x )\phi( x') \rangle_q + i\theta_c(\tau' -\tau) \langle \phi(x' )\phi( x) \rangle_q \nn\\
&=&  \theta_c(\tau -\tau')D_+ (x,x') + \theta_c(\tau' -\tau)D_-(x,x'),  \nn\\
\end{eqnarray}
where
\begin{eqnarray}
D_+ (x,x') &=& i\langle  \phi(x )\phi( x')\rangle_q, \nn\\
\end{eqnarray}
and
\begin{eqnarray}
D_- (x,x')  &=& i\langle  \phi(x' )\phi( x)\rangle_q
\end{eqnarray}
are the advanced and the retarded propagators respectively. $\hat{T}_c$ represents the time ordering operator and $\tau$ and $\tau'$ are   two arbitrary points on the contour C (Fig.~\ref{fig1}).

To investigate the region in the complex time plane (see ref.~\cite{Mills}) where $D_+(x,x')$ and $D_-(x,x')$ are defined, let us  consider the advanced part of the thermal propagator.

By introducing the completeness relation, $\sum_{m}^{}|m\rangle\langle m|$=\(  \mathbb{1} \), where $|m\rangle$ defines the spectrum of $H$, and by replacing the Heisenberg field $\phi(x)$ by Schrodinger field as $\phi(x)=e^{i\hat{H}\tau}  \phi(0,\vec{x})e^{-i\hat{H}\tau}$, we get,  

\begin{eqnarray*}
	D_+(x,x') =&&iZ_q^{-1}\sum_{m,n}^{ } \big[1+(q-1)\beta E_m\big]^{\frac{q}{1-q}} \langle m|e^{ {i\hat{H}\tau} } \phi(0,\vec{x})e^{ {-i\hat{H}\tau} }|n\rangle\langle n|e^{ {i\hat{H}\tau'} } \phi(0,\vec{x}')e^{ {-i\hat{H}\tau'} }|m\rangle 
	\\=&&iZ_q^{-1} \sum_{m,n}^{ } e^{ \frac{q}{1-q}\ln[1+(q-1)\beta E_m]} e^{ iE_m(\tau-\tau')} e^{ - iE_n(\tau-\tau') }\langle m|  \phi(0,\vec{x}) |n\rangle\langle n|  \phi(0,\vec{x}') |m\rangle 
	\\=&&iZ_q^{-1} \sum_{m,n}^{ } e^{iE_m\left(\tau-\tau'+\frac{i  \ln[1+(q-1)\beta E_m] }{(q-1)E_m}\right)}  e^{ - iE_n(\tau-\tau') }\langle m|  \phi(0,\vec{x}) |n\rangle\langle n|  \phi(0,\vec{x}') |m\rangle ,
\end{eqnarray*}
where we have used $\hat{H}|m\rangle=E_m|m\rangle$. The unbounded spectrum of $\hat{H}$ determines the convergence of the sums and consequently choose the domain in which $D_+(x,x')$ can be defined. The sum over $m$ and $n$ converge respectively for,
\begin{eqnarray*}
	&& \text{Im}  \left[\tau-\tau'+i\frac{  \ln[1+(q-1)\beta E_m] }{(q-1)E_m} \right]   \geq 0  
	\\&& \implies	\text{Im}( \tau-\tau')\geq  -\frac{   \ln[1+(q-1)\beta E_m] }{(q-1)E_m} ,
\end{eqnarray*}
and $\text{Im}( \tau-\tau')\leq 0$ . Combining these two we get, 
\begin{equation}
\label{7}
  -\frac{   \ln[1+(q-1)\beta E_m] }{(q-1)E_m} \leq \text{Im}(\tau-\tau')\leq 0.
\end{equation}
In the limit $q \ \rightarrow 1$, 
the above convergence condition becomes that of the Boltzmann-Gibbs statistics, 
\begin{equation}
 -\beta  \leq \text{Im}(\tau-\tau')\leq 0.
\end{equation}

In the case of Tsallis statistics, we have to consider a contour which always respects the convergence condition eq.~\eqref{7}. 
We can take the symmetrical contour suggested in Refs.  \cite{Schwinger,Keldysh,Niemi,Umezawa1,Umezawa2}
shown in Fig.~\ref{fig1}.  

\subsection{KMS relation in the Tsallis statistics}
\label{subsec:KMS}

We derive below the Kubo-Martin-Schwinger (KMS) relation (\cite{Kubo,Martin})
within the scope of Tsallis statistics. The KMS relation is the one connecting the retarded 
and the advanced parts of the propagator. Writing the explicit form of the retarded propagator we 
obtain the following equation,
\begin{eqnarray}
D_-(x,x') &=& i \langle \phi(x' )\phi( x) \rangle_q \nn\\
&=& iZ_q^{-1}\text{Tr}' \left[\left\{e^{-\beta\hat{H}}_q\right\}^q \phi(\tau',\vec{x}' )\phi( \tau,\vec{x})\right] \nn\\
&=& iZ_q^{-1}\text{Tr}' \left[\left\{e^{-\beta\hat{H}}_q\right\}^q \phi(\tau',\vec{x}' )\left\{e^{-\beta\hat{H}}_q\right\}^q 
\left\{e^{\beta\hat{H}}_{2-q}\right\}^q \phi( \tau,\vec{x})\right] \nn\\
&=& iZ_q^{-1}\text{Tr}' \left[\left\{e^{-\beta\hat{H}}_q\right\}^q \left\{e^{\beta\hat{H}}_{2-q}\right\}^q
\phi( \tau,\vec{x}) \left\{e^{-\beta\hat{H}}_q\right\}^q \phi(\tau',\vec{x}' )\right] \nn\\
&=& iZ_q^{-1}\text{Tr}' \left[\left\{e^{-\beta\hat{H}}_q\right\}^q e^{\frac{q}{q-1}\ln\left[1+(q-1)\beta \hat{H}\right]} e^{i\hat{H}\tau}  \phi(0,\vec{x})e^{-i\hat{H}\tau} e^{-\frac{q}{q-1}\ln[1+(q-1)\beta \hat{H}]} \phi(\tau',\vec{x}' )\right] \nn\\
&=& iZ_q^{-1}\text{Tr}' \left[\left\{e^{-\beta\hat{H}}_q\right\}^q  \phi\left(\tau-\frac{ i q}{E_m(q-1)} \ln \left[1+(q-1) \beta E_m\right] ,\vec{x} \right)    \phi(\tau',\vec{x}' ) \right] \nn\\
&=& D_+ \left(\vec{x},\vec{x}',\tau-\frac{ i q}{E_m(q-1)} \ln \left[1+(q-1)\beta E_m\right],\tau'\right) \nn\\
&=& D_+ \left(\vec{x},\vec{x}',\tau-i\bar{\beta},\tau'\right),
\label{16}     
\end{eqnarray}
where
\begin{equation}
\bar{\beta}=\frac{q}{(q-1)E_m} \ln\left[1+(q-1)\beta E_m\right],
\label{betabar}
\end{equation}
and 
\be
\left\{e^{-\beta\hat{H}}_q\right\}^q 
\left\{e^{\beta\hat{H}}_{2-q}\right\}^q =  \mathbb{1} .
\ee

It can be observed from eq.~\eqref{16} that the KMS relation acts as a boundary condition that imposes a periodicity in the propagator in terms of the time variable $\tau$. It is also observed that the KMS relation of the Tsallis statistics is energy dependent \cite{biro}. In the limit $q\rightarrow1$, eq.~\eqref{16} reduces to its Boltzmann-Gibbs counterpart \cite{bellac,das},
\begin{equation}
 D_-(x,x') =D_+(\vec{x},\vec{x}',\tau-  i  \beta  ,\tau'),
\end{equation}
and the energy dependence do not appear. It is worthwhile to mention that the periodicity condition is exclusive to the bosonic particles. For the fermions, one encounters anti-periodicity in the time axis (see section~\ref{sec:propfermion}).

The invariance of the two point functions 
under time translation stated as, $D_{+}(-\tau,0)= D_{-}(0,\tau)$ can be  ensured by the condition
${\cal{D}}_+(-\omega)={\cal{D}}_-(\omega)$, where ${\cal{D}}_+$ and ${\cal{D}}_-$ are the Fourier transforms 
of $D_+(-\tau,0)$ and $D_-(0,\tau)$ respectively. 
It may be noted here that in the limit $q\rightarrow 1$,
$\bar{\beta}\rightarrow \beta$ and consequently it becomes straightforward  to show the time invariance of the two point functions. 
\\
\\
\subsection{Calculating Tsallis propagator using the differential equation method}

After obtaining the KMS relation, we derive the thermal propagator for the real scalar particles
by solving the corresponding field equation, {\it i.e.} we need to solve the following equation,
\begin{equation}
(\scalebox{1.1}{$\square$}_{c}+m^2)D(x,x')=\delta^4_c(x-x').
\label{scalarEOM}
\end{equation}
We can take the following Fourier transform as the spatial coordinates are not associated with boundary conditions as opposed to the time coordinate:

\begin{equation}
 D(x,x')=\int \frac{d^3\bold{k}}{(2\pi)^3}e^{-i\vec{k}.(\vec{x}-\vec{x}')}D(\vec{k},\tau,\tau').
 \label{propfourier}
\end{equation}
Putting eq.~\eqref{propfourier} in eq.~\eqref{scalarEOM} and using the integral representation of the space-dependent part of the Dirac-delta function, we obtain
\begin{eqnarray}
\int \frac{d^3\bold{k}}{(2\pi)^3} \left(\frac{\partial^2}{\partial\tau^2}+\bold{k}^2+m^2\right) e^{-i\vec{k}.(\vec{x}-\vec{x}')}D(\vec{k},\tau,\tau')
= \delta_c (\tau-\tau') \int \frac{d^3k}{(2\pi)^3}e^{-i\vec{k}.(\vec{x}-\vec{x}')} . \nn\\
\end{eqnarray}
Comparing the r.h.s with the l.h.s we get,
\begin{equation}
\left(\frac{\partial^2}{\partial\tau^2}+\omega_{\bold{k}}^2\right)D(\vec{k},\tau,\tau')=\delta_c (\tau-\tau'),
\label{scalarEOMmomsp}
\end{equation}
where $\omega_{\bold{k}}=\sqrt{\bold{k}^2 +m^2}$ is the single particle energy, and $\bold{k} \equiv |\vec{k}|$.

For $\tau \neq \tau'$, the solution of eq.~\eqref{scalarEOMmomsp} is a linear combination of $e^{i \omega_{\bold{k}} \tau}$ and $e^{-i \omega_{\bold{k}} \tau}$. We can write the retarded and the advanced part of the 
Green's function (or propagator) as the linear combinations of these functions with coefficients depending on $\tau'$,
\begin{eqnarray}
 D_+(\vec{k},\tau,\tau')&=& A_1(\tau')e^{- i \omega_{\bold{k}} \tau}+A_2(\tau')e^{  i \omega_{\bold{k}} \tau} ~  (\tau >\tau'), \nn\\
 D_-(\vec{k},\tau,\tau')&=&B_1(\tau')e^{- i \omega_{\bold{k}} \tau}+B_2(\tau')e^{  i \omega_{\bold{k}} \tau} ~ (\tau < \tau'). \nn\\
 \label{23} 
\end{eqnarray}
The unknown coefficients can be found with the help of the modified KMS relation in eq.~\eqref{16} as one of the boundary conditions along with the following ones \cite{arfken},
\begin{eqnarray}
D_+(\vec{k},\tau,\tau') &=& D_-(\vec{k},\tau,\tau') \\
\frac{\partial D_+(\vec{k},\tau,\tau')}{\partial\tau} &-& \frac{\partial D_-(\vec{k},\tau,\tau')}{\partial\tau} =1.
\label{BCGreen}
\end{eqnarray}
Using the above boundary conditions in eq.~\eqref{23} we get
\begin{equation}
\label{26}
B_1(\tau')=\Big [A_1(\tau')e^{-i\omega_{\bold{k}}\tau' }+\frac{1}{2i\omega_{\bold{k}}}\Big ] e^{ i\omega_{\bold{k}}\tau' },
\end{equation}
and 
\begin{equation}
\label{27}
B_2(\tau')=\Big [A_2(\tau')e^{ i\omega_{\bold{k}}\tau' }-\frac{1}{2i\omega_{\bold{k}}}\Big ] e^{ -i\omega_{\bold{k}}\tau' },
\end{equation}
which when put into eq.~\eqref{23}, yield,

\begin{eqnarray}
D_-(\vec{k},\tau,\tau')
  = \frac{e^{ i\omega_{\bold{k}}(\tau'-\tau) }- e^{ -i\omega_{\bold{k}}(\tau'-\tau) }}{2i\omega_{\bold{k}}}+D_+(\vec{k},\tau,\tau').
  \label{scaladretrel}
\end{eqnarray}
Using eqs.~\eqref{23} and \eqref{16},  we get
 \begin{eqnarray}
 A_1(\tau')e^{-i\omega_{\bold{k}}(\tau-i\bar{\beta})}
 + 
 A_2(\tau')e^{-i\omega_{\bold{k}}(\tau-i\bar{\beta})}
& =& \frac{e^{ i\omega_{\bold{k}}(\tau'-\tau) }- e^{ -i\omega_{\bold{k}}(\tau'-\tau) }}{2i\omega_{\bold{k}}}
\nn\\
 &&
 +A_1(\tau')e^{-i\omega_{\bold{k}}\tau  }+A_2(\tau')e^{ i\omega_{\bold{k}}\tau  } . \nn\\
 \label{30}
 \end{eqnarray}      
                                 
 Now, $E_m$ is the total energy of the state $|m>$' which is populated by many particles. If there are $n_{k'}$ particles each with energy $\omega_{\bold{k}'}$, then we may write 
 \bea
 E_m=\sum_{k'} n_{k'} \omega_{\bold{k}'}.
 \eea
Comparing the coefficients of $e^{-i\omega_{\bold{k}}\tau  }$ and $e^{ i\omega_{\bold{k}}\tau  }$ in both the sides of eq.~\eqref{30} we get,

\begin{eqnarray}
\label{A1A2}
 A_1(\tau') &=& \frac{ \frac{-i}{2\omega_{\bold{k}}}e^{ i\omega_{\bold{k}}\tau'  }}{\Big[1+(q-1)\beta \sum n_{k'}\omega_{\bold{k}'} \Big]^{\frac{q\omega_{\bold{k}}} {(1-q)\sum n_{k'}\omega_{\bold{k}'}} }-1}, \nn\\
A_2(\tau') &=&\frac{ \frac{i}{2\omega_{\bold{k}}}e^{-i\omega_{\bold{k}}\tau'}} {\Big[1+(q-1)\beta \sum n_{k'}\omega_{\bold{k}'} \Big]^{ \frac{ q\omega_{\bold{k}}} {(q-1)\sum n_{k'}\omega_{\bold{k}'}} }-1}. \nn\\
\end{eqnarray} 
At this point, we introduce the following approximation for the power-law,   
\begin{eqnarray}
\Big[1+(q-1)\beta \sum n_{k'}\omega_{\bold{k}'} \Big]^{ \frac{ q\omega_{\bold{k}}} {(q-1)\sum n_{k'}\omega_{\bold{k}'}} } \nn\\
\approx
\Big[1+(q-1)\beta \omega_{\bold{k}}\Big]^{ \frac{q}{q-1} \frac{\cancel{\sum n_{k'}\omega_{\bold{k}'}}} 
{\cancel{\sum n_{k'}\omega_{\bold{k}'}}} }.
\label{factapprox}
\end{eqnarray} 
If we write,
\begin{eqnarray}
n_{\text{T}}(\omega_{\bold{k}})=\frac{1}{\left[1+(q-1)\beta \omega_{\bold{k}})\right]^{\frac{q}{q-1}}-1}, \nn\\
\end{eqnarray} 
$A_1(\tau')$, and $A_2(\tau')$ can be written as, 
\begin{eqnarray}
A_1(\tau') &= &\frac{ ie^{ i\omega_{\bold{k}}\tau'  }}{2\omega_{\bold{k}}} \big[1+n_ {\text{T}}(\omega_{\bold{k}})\big],
\\ 
A_2(\tau') &=& \frac{ ie^{-i\omega_{\bold{k}}\tau'  }}{2\omega_{\bold{k}}} n_ {\text{T}}(\omega_{\bold{k}}).
\end{eqnarray}

\noindent Using the values of $A_1(\tau')$and $A_2(\tau')$, we get $B_1(\tau')$and $B_2(\tau')$,
\begin{eqnarray}
B_1(\tau') &=& \frac{ ie^{i\omega_{\bold{k}}\tau'  }}{2\omega_{\bold{k}}} n_ {\text{T}}(\omega_{\bold{k}}), \\
B_2(\tau') &=& \frac{ ie^{-i\omega_{\bold{k}}\tau'  }}{2\omega_{\bold{k}}} \big[1+n_ {\text{T}}(\omega_{\bold{k}})\big].
\end{eqnarray}

\noindent Hence, the thermal propagator can be written as,
%
\begin{eqnarray}
D(\vec{k},\tau_i,\tau'_j) &\approx& \frac{ i}{2\omega_{\bold{k}}} \Big [\theta_c(\tau_i  -\tau_j' ) \Big \{ (1+n_{\text{T}})  e^{ -i\omega_{\bold{k}}(\tau_i  -\tau_j' )} + n_{\text{T}} e^{  i\omega_{\bold{k}}(\tau_i  -\tau_j' )}\Big  \} \nn\\
&&+ \theta_c(\tau_j'  -\tau_i  ) \Big\{ n_{\text{T}}  e^{ -i\omega_{\bold{k}}(\tau_i  -\tau_j'  )} +(1+ n_{\text{T}})  e^{  i\omega_{\bold{k}}(\tau_i  -\tau_j' )}\Big \}\Big ], 
\label{42}
\end{eqnarray}
where the indices $i$, and $j$ denote on which part (1 or 2) of the contour $\tau$ lies (see Fig.~\ref{fig1}).

The components of the momentum space  thermal propagator can  be found by taking temporal Fourier transform
\begin{equation}
D_{ij}(\vec{k},k_0)=\int_{-\infty}^{\infty}dt e^{ik_0(t-t')}D(\vec{k},\tau_i,\tau'_j).
\label{43}
\end{equation}

\noindent Now, the above integration in eq.~\eqref{43} can be performed by choosing two points on any one of the two horizontal line
shown in Fig.~\ref{fig1}. Propagators containing $\tau$ lying on the vertical lines vanish due to the Riemann-Lebesgue  lemma \cite{Titchmarsh}. We can take two points on two horizontal lines in four different ways,
\begin{itemize}
\item $\tau$ and $\tau'$ both lie on line 1.
\item $\tau$ and $\tau'$ both lie on line 2.
\item $\tau$ lies on line 1 and $\tau'$ lies on line 2.
\item $\tau$ lies on line 2 and $\tau'$ lies on line 1 .
\end{itemize}
These four cases can be implemented through the following choices of theta functions in that order,
\begin{enumerate}
\item  $\theta_c(\tau_1-\tau'_1)=\theta(t-t')$.
\item  $\theta_c(\tau_1-\tau'_2)=0$.
\item  $\theta_c(\tau_2-\tau'_1)=1$.
\item  $\theta_c(\tau_2-\tau'_2)=\theta(t'-t )$.
\end{enumerate}

Let us now calculate the `11' component of the real time thermal propagator which may be written as follows,
\begin{eqnarray}
D_{11}(\vec{k},k_0)&=&\frac{ i}{2\omega_{\bold{k}}}\int_{-\infty}^{\infty}dt ~e^{ik_0(t-t')}\Big [\theta (t  -t' ) \Big \{(1+n_{\text{T}})  e^{ -i\omega_{\bold{k}}(t  -t' )} + n_{\text{T}}  e^{  i\omega_{\bold{k}}(t  -t' )}\Big \} 
\nn\\ 
&&+ \theta (t'  -t  )\Big \{  n_{\text{T}}  e^{ -i\omega_{\bold{k}}(t  -t' )} +(1+ n_{\text{T}})  e^{  i\omega_{\bold{k}}(t  -t' )} \Big \}\Big ] 
\nn\\ 
&=& \frac{ i}{2\omega_{\bold{k}}} \int_{-\infty}^{\infty}dt ~e^{ik_0t}\Big [\theta (t) \Big \{(1+n_{\text{T}})  e^{ -i\omega_{\bold{k}} t } + n_{\text{T}}  e^{  i\omega_{\bold{k}} t }\Big \} 
\nn\\
&&
+ \theta (-t)\Big \{  n_{\text{T}}  e^{ -i\omega_{\bold{k}}t} +(1+ n_{\text{T}}) e^{  i\omega_{\bold{k}} t } \Big \}\Big ]  \nn\\
&=& \frac{ i}{2\omega_{\bold{k}}} \int_{0}^{\infty}dt ~\Big [(1+n_{\text{T}})  e^{ i(k_0-\omega_{\bold{k}}) t } + n_{\text{T}}  e^{  i(k_0+\omega_{\bold{k}}) t } 
\nn\\
&&
+ n_{\text{T}}  e^{ -i(k_0-\omega_{\bold{k}})t} +(1+ n_{\text{T}}) e^{  -i(k_0-\omega_{\bold{k}}) t } \Big ]  \nn\\
&=& \lim_{\epsilon \rightarrow 0} \frac{ i}{2\omega_{\bold{k}}} \int_{0}^{\infty}dt  \Big [ (1+n_{\text{T}})  e^{  i(k_0-\omega_{\bold{k}}+i\epsilon)  t } + n_{\text{T}}  e^{  i(k_0+\omega_{\bold{k}}+i\epsilon)t } \nn\\
 &&+   n_{\text{T}}  e^{  -i(k_0-\omega_{\bold{k}}-i\epsilon)t} +(1+ n_{\text{T}}) e^{  -i(k_0+\omega_{\bold{k}}-i\epsilon) t } \Big ].  \nn\\
 \end{eqnarray}
where we have replaced $(t-t')$ by $ t$ which is again replaced with $-t$ in the second term involving $\theta(-t)$.
Now, the integrands are oscillatory and they diverge at $t \rightarrow \infty$. To avoid this divergence while carrying out
the integral, we have added an infinitesimally small imaginary part in $k_0$ which will be made to vanish finally.
After integrating, we obtain,
\begin{eqnarray}
D_{11}(\vec{k},k_0)= \frac{n_{\text{T}}(\omega_{\bold{k}})}{k_0^2-\omega_{\bold{k}}^2-i\epsilon} - 
\frac{1+n_{\text{T}}(\omega_{\bold{k}})}{k_0^2-\omega_{\bold{k}}^2+i\epsilon}
\end{eqnarray}
We simplify the above equation further by using the  Sokhotski-Plemelj formula \cite{plemeljform},
\begin{equation}
\lim\limits_{\epsilon\rightarrow 0}\frac{1}{x \pm  i\epsilon}=\mathcal{P}\left (\frac{1}{x}\right)\mp i \pi \delta(x),
\end{equation}
where `$\mathcal{P}$' denotes the principal value, to obtain,
\begin{equation}
 D_{11}(\vec{k},k_0)=\frac{-1}{k^2-m^2+i\epsilon}+\frac{2 \pi i \delta(k^2-m^2 )} {\left[1+(q-1)\beta \omega_{\bold{k}}\right]^{\frac{q}{q-1}}-1},
 \label{D11TBEfinal}  
\end{equation}
where  $k_0^2-\omega_{\bold{k}}^2=k^2-m^2$, for invariant mass $m$.
In the limit $q\rightarrow 1$, we get back the real time BG bosonic propagator,
\begin{equation}
D^{\text{BG}}_{11}(\vec{k},k_0)=\frac{-1}{k^2-m^2+i\epsilon}+ {\frac{2 \pi i   \delta(k^2-m^2 ) }{e^{\beta \omega_{\bold{k}}} -1}}. 
\label{D11BE}
\end{equation}
If we compare with the BG thermal field theoretic results in eq.~\eqref{D11BE}, the coefficient of the quantity $2\pi i \delta(k^2-m^2)$ in eq.~\eqref{D11TBEfinal} can be identified to be the Tsallis Bose-Einstein distribution given by,
\begin{equation}
n^{\text{BE}}_{\text{T}} (\omega_{\bold{k}}) \equiv n_{\text{T}} (\omega_{\bold{k}}) = \frac{1 }{\left[1+(q-1)\beta \omega_{\bold{k}}\right]^{\frac{q}{q-1}}-1}=
\frac{1}{e^{\bar{\beta}\omega_k}-1}.
\end{equation}
where 
\begin{equation}
\bar{\beta}(x)=\frac{q}{(q-1)x}ln[1+(q-1)\beta x]
\end{equation}
It can easily be checked that in the limit, $q\rightarrow 1$ one gets $\bar{\beta}\rightarrow \beta$. It is interesting to note that
in this limit the TB statistics goes to GB statistics as a consequence it becomes simple to prove the invariance of two point functions 
under time translation.
The form of this distribution is identical with the result obtained in \cite{Hasegawa}. This result may also be obtained from
the factorization approximation of the zeroth term approximation of the result obtained in \cite{Parvan19}. The above two 
references calculate the Tsallis Bose-Einstein distribution from Tsallis statistical mechanical formulations, and
hence the form of the distribution obtained in eq.~\eqref{D11TBEfinal} belongs to the Tsallis statistical mechanics. 

Next we proceed to calculate the $D_{12}(\vec{k},k_0)$. 
In this case $\tau_1=t, \tau_2'=t'-i \bar{\beta}/2$ (see Fig.~\ref{fig1}) and $\theta(\tau_1-\tau_2')=0,\theta(\tau'_2-\tau_1)=1 $
\begin{eqnarray}
D_{12}(\vec{k},k_0)&\approx&\frac{ i}{2\omega_{\bold{k}}} 
\int_{-\infty}^{\infty}dt e^{ik_0(t-t')}\Big [\theta (\tau_1-\tau_2') \Big \{(1+n_{\text{T}}) e^{ -i\omega_{\bold{k}} (\tau_1-\tau_2')} + n_{\text{T}}^q  e^{  i\omega_{\bold{k}} (\tau_1-\tau_2')}\Big \} 
\nn\\ 
&&+\theta  (\tau'_2-\tau_1 )\Big \{  n_{\text{T}}  e^{ -i\omega_{\bold{k}}(\tau_1-\tau_2')} +(1+ n_{\text{T}})  e^{  i\omega_{\bold{k}}(\tau_1-\tau_2')} \Big \}\Big ] 
\nn\\ 
&=&\frac{ i}{2\omega_{\bold{k}}} 
\int_{-\infty}^{\infty}dt ~e^{ik_0 (t-t')}  \Big \{ n_{\text{T}}  e^{ -i \omega_{\bold{k}} \left(t-t'+ i\frac{\bar{\beta}}{2} \right) }
+ ( 1+n_{\text{T}})  e^{  i\omega_{\bold{k}} \left(t-t'+ i\frac{\bar{\beta}}{2} \right) } \Big \} 
\end{eqnarray}
Substituting $t-t'=t$ and using the integral representation of delta function 
\begin{equation}
\delta (x)=\frac{1}{2\pi}\int_{-\infty}^{\infty}dk ~e^{\pm i kx},
\end{equation}
and using the approximation eq.~\eqref{factapprox}, we finally get,
\begin{equation}
 D_{12}(\vec{k},k_0) \approx 2 \pi i \sqrt{n_{\text{T}} (1+n_{\text{T}}) }  \delta(k^2-m^2).
\end{equation}
In the limit $q\rightarrow 1$, we get back the usual Boltzmann-Gibbs propagator,
\begin{equation}
D_{12}(\vec{k},k_0)= 2 \pi i \sqrt{n( 1+n )}\delta(k^2-m^2),
\end{equation}
where $n$ is the Boltzmann-Gibbs bosonic distribution.
Similarly we get, $D_{21}(\vec{k},k_0)=D_{12}(\vec{k},k_0) $ and $D_{22}(\vec{k},k_0)=-D^*_{11}  (\vec{k},k_0) $.  
Then the matrix propagator ${\bf D}(\vec{k},k_0)$ is given by,  
\begin{equation}
{\bf D}(\vec{k},k_0)=  \def\arraystretch{2 }
\begin{pmatrix}
\triangle_{\text{F}}+2 \pi i n_ {\text{T}} \delta(k^2-m^2 )  
& 
2 \pi i \sqrt{n_{\text{T}} ( 1+n_{\text{T}})}  \delta(k^2-m^2)  
\\
2 \pi i \sqrt{n_{\text{T}} ( 1+n_{\text{T}})}  \delta(k^2-m^2) 
& 
-\triangle_{\text{F}}^*+2 \pi i n_ {\text{T}} \delta(k^2-m^2 )
\\
\end{pmatrix}  ,
\label{propmatrix}
\end{equation}
where $\Delta_{\text{F}}$ is the zero temperature term in the propagator.
Substituting $\Delta_{\text{F}}-\Delta_{\text{F}}^*$ in place of $2\pi i \delta (k^2-m^2)$the propagator 
in eq.~\eqref{propmatrix} can be put in a diagonal form,
\begin{equation}
\label{2.50}
{\bf D}(\vec{k},k_0)={\bf U}(k_0) \def\arraystretch{1 }
\begin{pmatrix}
\triangle_F  & 0\\
  0 & -\triangle_F^*  \\
\end{pmatrix} {\bf U}(k_0).
\end{equation}
The diagonalizing matrix ${\bf U}$ is given by,
\begin{equation}
\label{2.51}
{\bf U}(k_0)
=\def\arraystretch{2.5 }
\begin{pmatrix}
\sqrt{1+n_{\text{T}}}
&  
\sqrt{n_{\text{T}}}
\\
\sqrt{n_{\text{T}}}
& 
\sqrt{1+n_{\text{T}}},
\\
\end{pmatrix},
\end{equation}
Now, for free propagator
 \bea
 D_{11}&=&\triangle_F+\left(\triangle_F-\triangle_F^*\right)n_{\text{T}} \nn\\
&=& \left(1+n_{\text{T}} \right) \triangle_F - n_{\text{T}} \triangle_F^*,
 \eea
which implies,
\bea
\text{Re}D_{11}+i\text{Im}D_{11}&=& \text{Re} \triangle_F+i\text{Im}\triangle_F
(1+2n_{\text{T}}). \nn\\
 \eea
Comparing the real and the imaginary parts from both the sides we obtain,
\bea
\text{Re}D_{11}(\vec{k},k_0) &=& \text{Re} \triangle_F,~\text{and} \nn\\
\text{Im}D_{11}(\vec{k},k_0)  &=& \epsilon(k_0)\coth \left[\frac{q}{2(q-1)}\ln\left[1+\beta(q-1)k_0\right]\right]
\text{Im}\triangle_F, \nn\\
\label{relation}
\eea
where the sign function $\epsilon(k_0)$ is used to carry out the following transformation,
\bea
\coth \left[\frac{q}{2(q-1)}\ln \big[1+\beta(q-1)|k_0|\big] \right] =\epsilon(k_0)\coth \left[\frac{q}{2(q-1)}\ln\left[1+\beta(q-1)k_0\right]\right].\nn\\
\eea


\section{Free Tsallis thermal propagator for the fermions}
\label{sec:propfermion} 
The Dirac thermal propagator in the Tsallis statistics may be defined by,
\begin{eqnarray}
S(x,x')=&&i\langle T_c\psi(x)\bar{\psi}(x')\rangle_q
\\=&&iZ_q^{-1}Tr[\rho^q T_c\psi(x)\bar{\psi}(x')].
\end{eqnarray}
where 
$Z_q=\text{Tr}[\left\{\hat{\rho}\right\}^q]$, is the partition function for vanishing chemical potential. 
The Dirac thermal propagator can be written as, 
\begin{eqnarray}
S(x,x') &=& i\theta_c(\tau -\tau') \langle  \psi(\vec{x},\tau )\bar{\psi}(\vec{x}',\tau' )\rangle_q \nn\\
&&- i\theta_c(\tau' -\tau ) \langle  \psi(\vec{x}',\tau' )\bar{\psi}(\vec{x},\tau )\rangle_q \nn\\
&=&\theta_c(\tau -\tau') S_+(x,x')+\theta_c(\tau' -\tau )S_-(x,x'), \nn\\
\end{eqnarray}

where
\begin{eqnarray}
S_+(x,x')=&&i  \langle  \psi(\vec{x},\tau )\bar{\psi}(\vec{x}',\tau' )\rangle_q,
\\S_-(x,x')=&&-i \langle  \psi(\vec{x}',\tau' )\bar{\psi}(\vec{x},\tau )\rangle _q.
\end{eqnarray}

Next, we quote the KMS relation for the Dirac field. The derivation is similar to that for the bosons.
\begin{eqnarray}
S_-(x,x') =
- S_+(\vec{x},\vec{x}',\tau-i\bar{\beta},\tau'),  
\label{fermKMS}   
\end{eqnarray}
where $\bar{\beta}$ is given by eq.~\eqref{betabar}. In the limit $q \ \rightarrow 1$, the equation reduces to its
Boltzmann-Gibbs counterpart,
\begin{equation}
S_-(x,x') =-S_+(\vec{x},\vec{x}',\tau-  i  \beta  ,\tau').
\end{equation}

Calculations in the scalar case and in the Dirac case closely follow each other if we write the fermion propagator in terms of the scalar propagator in the following manner, 
\begin{equation}
S(x,x')=(i \slashed{\partial}_c  +m)D(x,x').
\label{SfnD}
\end{equation}
Spatial Fourier transform of $S(x,x')$ yields,
\begin{equation}
 S(x,x')=\int_{-\infty}^{\infty}\frac{d^3p}{(2\pi)^3}e^{i\vec{p}\cdot(\vec{x}-\vec{x}')}S(\vec{p},\tau,\tau').
 \label{SFT}
\end{equation}
With the help of eqs.~\eqref{SfnD}, and \eqref{SFT} we obtain,
\begin{equation}
S(\vec{p},\tau,\tau') =
(i\gamma^0 \partial_0-\vec{\gamma}\cdot{\vec{p}}+m) D(\vec{p},\tau,\tau').
\end{equation}
The advances and retarded parts are given by,
\begin{eqnarray}
 S_+(\vec{p},\tau,\tau')=&&(i\gamma^0 \partial_0-\vec{\gamma}\cdot{\vec{p}}+m) D_+(\vec{p},\tau,\tau')
 \\ S_-(\vec{p},\tau,\tau')=&&(i\gamma^0 \partial_0-\vec{\gamma}\cdot{\vec{p}}+m) D_-(\vec{p},\tau,\tau'),
\end{eqnarray}
Guided by eqs.~\eqref{23}, \eqref{scaladretrel} from the scalar case and by taking the spatial Fourier transform of the 
eq.~\eqref{fermKMS} we obtain the following equation, 
%

%
%
%
%
%
%
%
\begin{eqnarray}
   S_+(\vec{p},\tau,\tau')- \frac{i}{2\omega}\{ C_+ e^{-i\omega(\tau -\tau ')}-C_- e^{ i\omega(\tau -\tau ')}\}  =- S_+ \left(\vec{p},\tau -
   i\bar{\beta},\tau'\right) , 
\end{eqnarray}
where $C_{\pm}=(\pm\gamma^0\omega-\vec{\gamma}.\vec{p}+m)$ and $\omega=\sqrt{|\vec{p}|^2+m^2}$.
Equating the coefficients of $e^{ \pm i \omega \tau   }$, and employing the approximation indicated in 
eq.~\eqref{factapprox} 
we get,
\begin{eqnarray}
 A^\prime_1(\tau') &=& \frac{i}{2\omega} \frac{e^{  i\omega \tau'   }}{1+e^{-\frac{q}{(q-1) } \ln[1+(q-1)\beta ]}}, \nn\\
 A^\prime_2(\tau') &=& \frac{-i}{2\omega_{\bold{k}}}\frac{e^{ -i\omega \tau'   }}{1+e^{\frac{ q}{(q-1) } \ln[1+(q-1)\beta \omega ]}}.
\end{eqnarray}
%
Identifying 
\begin{equation}
f_{\text{T}}(\omega)=\frac{1}{[1+(q-1)\beta \omega]^\frac{q}{q-1}+1}=\frac{1}{e^{\bar{\beta}\omega}+1},
\label{FD}
\end{equation}
which is the single particle distribution for the Tsallis fermions,
we can write $A^\prime_1$, and $A^\prime_2$ as,

\begin{eqnarray}
A^\prime_1(\tau')=&&\frac{  ie^{  i\omega \tau'}}{2\omega} 
(1-f_T),
\\ A^\prime_2(\tau')=&&\frac{ - ie^{  -i\omega \tau'}}{2\omega} {} f_{\text{T}}.
\end{eqnarray}

So, the fermion thermal propagator can be written as,
\begin{eqnarray}
 S(\vec{p},\tau,\tau')  &=& \theta_c(\tau -\tau') S_+(\vec{p},\tau,\tau')+\theta_c(\tau' -\tau )S_-(\vec{p},\tau,\tau')
 \nn\\ 
 &=&
 \theta_c(\tau -\tau') S_+(\vec{p},\tau,\tau')+\theta_c(\tau' -\tau ) S_+(\vec{p},\tau,\tau')\nn\\
 &&-\frac{  i}{2\omega}\theta_c(\tau' -\tau )\{ C_+ e^{-i\omega_{\bold{k}}(\tau -\tau ')}-C_- e^{ i\omega (\tau -\tau ')}\} \nn\\
 &=& 
 S_+(\vec{p},\tau,\tau') -\frac{  i}{2\omega }\theta_c(\tau' -\tau )\{ C_+ e^{-i\omega (\tau -\tau ')}-C_- e^{ i\omega (\tau -\tau ')}\} 
 \nn\\
 &=& 
 \frac{  i}{2\omega }    \Big [e^{- i\omega (\tau -\tau ')}C_+ \Big \{(1-f_{\text{T}}) -\theta_c(\tau' -\tau )\Big\}-e^{  i\omega
 (\tau -\tau ')}C_-\Big\{f_{\text{T}} -\theta_c(\tau' -\tau )\Big\} \Big ] 
 \nn\\
 &=& \frac{  i}{2\omega }    \Big [e^{- i\omega (\tau -\tau ')}C_+ \Big \{(\theta_c(\tau -\tau' )-f_{\text{T}}) \Big\}+e^{  i\omega (\tau -\tau ')}C_-\Big\{ \theta_c(\tau' -\tau ) -f_{\text{T}}\Big\} \Big ] . \nn\\
\end{eqnarray}
Taking temporal Fourier transform, and invoking the time translational invariance,
\begin{eqnarray}
 S_{ij}(\vec{p},p_0)&=& \int_{-\infty}^{\infty}dt ~e^{ip_0(t-t')}S(\vec{p}, \tau_i-\tau'_j) 
 \nn\\
 &=& \frac{  i}{2\omega}\int_{-\infty}^{\infty}dt ~e^{ip_0(t-t')}
 \Bigg [e^{- i\omega (\tau_i -\tau_j ')}C_+  \Big\{\theta_c(\tau_i -\tau_j' )-f_{\text{T}} \Big\}
\nn\\
&&
 +e^{  i\omega (\tau_i -\tau _j')}C_-\Big\{\theta_c(\tau_j' -\tau_i )-f_{\text{T}}\Big \} \Bigg ],
\end{eqnarray}
where, as usual, $i,~j$ represent points on the lines on the contour $C$. 

The computation of the $i=1,~j=1$ components of $S$ yields,
\begin{eqnarray}
S_{11}(\vec{p},p_0) &=& \frac{  i}{2\omega} \int_{-\infty}^{\infty}dt ~e^{ip_0(t-t')}
\Bigg [e^{- i\omega (t-t')}C_+  
\Big\{\theta (t-t' )-f_{\text{T}}\Big\} 
\nn\\
&&
+e^{  i\omega (t-t')} C_-\Big\{\theta (t'-t )-f_{\text{T}}\Big \} \Bigg ]
\nn\\
&=&-\frac{\pi i}{\omega} f_{\text{T}} \Big[C_+\delta(p_0-\omega)+C_-\delta(p_0+\omega)\Big]
+\frac{i}{2\omega} \left [\frac{C_-}{i(p_0+\omega-i\epsilon)}-\frac{C_+}{i(p_0-\omega+i\epsilon)}\right]
\nn\\
&=& (\slashed{p}+m) \triangle_F(p)- \frac{2\pi i(\slashed{p}+m)\delta(p^2-m^2)}{[1+(q-1)\beta \omega]^\frac{q}{q-1}+1},
\label{Fermi}
\end{eqnarray}
where $\slashed{p}=\gamma^0p_0-\vec{\gamma}.\vec{p}$. Also, we use the identity $\theta (t'-t )+\theta (t -t' )=1 $, 
replace $(t-t')\rightarrow t$, and utilize the integral representation of the Dirac-delta function.
%
Comparing the expression of $S_{11}$ in eq.~\eqref{Fermi} with the BG field theoretic results, we can identify $f_{\text{T}}$
given in eq.~\eqref{FD} as the Tsallis distribution for the Dirac particles.
Next we compute $S_{12}$,
\begin{eqnarray}
 S_{12}(\vec{p},p_0) &=& \int_{-\infty}^{\infty}dt ~e^{ip_0(t-t')}S(\vec{p}, \tau_1-\tau'_2) 
 \nn\\
 &=& \frac{i}{2\omega} 
 \int_{-\infty}^{\infty}dt ~e^{ip_0(t-t')}
 \Big[  -C_+ e^{-i\omega\left(t-t'+i\frac{\bar{\beta}}{2}\right)} f_T + C_- e^{ i\omega\left(t-t'+ i\frac{\bar{\beta}}{2} \right)} \left(1-f_{\text{T}}\right)\Big]
 \nn\\
& \approx & \frac{  \pi i}{\omega}\sqrt{f_{\text{T}}(1-f_{\text{T}})} \big[C_-\delta(p_0+\omega)
 -C_+\delta(p_0-\omega)\big].
\nn\\
  \end{eqnarray}
Let us define two quantities
\begin{equation}
 N_1=\sqrt{f_{\text{T}}} \big[ \theta(p_0)+ \theta(-p_0) \big],
\end{equation}
and 
\begin{equation}
N_2=\sqrt{(1-f_{\text{T}})} \big[ \theta(p_0)- \theta(-p_0) \big].
\end{equation}
Using which, we can write,
\begin{eqnarray}
S_{1 1}(\vec{p},p_0)=&&(\slashed{p}+m) \Big [\triangle_F(p)-2\pi i N_1^2 \delta(p^2+m^2)\Big ]
\nn\\
S_{1 2}(\vec{p},p_0)=&&-2\pi i(\slashed{p}+m)   N_1N_2 \delta(p^2-m^2).
\end{eqnarray}
Similarly we can get $S_{2 1}=-S_{1 2} $ and $S_{2 2}=-S_{1 1}^*$.
%
%
Hence, the matrix representing the real time propagator ${\bf S}(\vec{p},p_0)$ is given by,
\begin{equation}
{\bf S}(\vec{p},p_0)=  \def\arraystretch{2 }
(\slashed{p}+m)\begin{pmatrix}
 \triangle_F(p)-2\pi i N_1^2 \delta(p^2+m^2)  &  -2\pi i  N_1N_2 \delta(p^2+m^2)\\  2\pi i    N_1N_2 \delta(p^2+m^2) &  
 -\triangle^*_F(p)+2\pi i N_1^2 \delta(p^2+m^2)  \\
\end{pmatrix}  .
\end{equation}
Using $2\pi i\delta(p^2+m^2)=\triangle _F(p)-\triangle^*_F(p)$ and $ N_1^2+N_2^2=1$, the above matrix 
can be written as,
\begin{equation}
{\bf S}(\vec{p},p_0)=  \def\arraystretch{2 }
(\slashed{p}+m)\begin{pmatrix}
\triangle_F(p)-  N_1^2 \{\triangle _F(p)-\triangle^*_F(p)\}  &     N_1N_2 \Big[\triangle _F(p)-\triangle^*_F(p)\Big]\\    N_1N_2 \Big[\triangle _F(p)-\triangle^*_F(p)\Big] &  -\triangle^*_F(p)-  N_1^2 \{\triangle _F(p)-\triangle^*_F(p)\}   \\
\end{pmatrix}  .
\end{equation}

The above expression can be put in a diagonal form as follows, 

\begin{equation}
\label{3.27}
{\bf S} (\vec{p},p_0)=(\slashed{p}+m)\bf{V}  \def\arraystretch{1 }
\begin{pmatrix}
\triangle_F  & 0\\
0 & - \triangle_F^*  \\
\end{pmatrix}\bf{V}  ,
\end{equation}

where the diagonalizing matrix $\bf{V}$ is given by,

\begin{equation}
\label{3.28}
\bf{V}
=\def\arraystretch{1.5 }
\begin{pmatrix}
 N_2    & - N_1  \\ N_1 &N_2\\
\end{pmatrix} .
\end{equation}
 
 We   find out the relationship between the `11' component of the free fermion propagator,
 and the free propagator(diagonalized) in a way similar to what has been done for the bosonic fields in
 eq.~\eqref{relation}. We simply quote the results below,
 \bea
\text{Re}S_{11}(\vec{p},p_0) &=& \text{Re} \left[(\slashed{p}+m)\triangle_F\right],~\text{and} \nn\\
\text{Im}S_{11}(\vec{p},p_0) &=& \epsilon(p_0)\tanh \left[\frac{q}{2(q-1)}\ln\left[1+\beta(q-1)p_0\right]\right]
\text{Im}\left[(\slashed{p}+m)\triangle_F\right]. \nn\\
\label{relationdirac}
\eea

\section{Diagonalization}

In the above calculations, we observe that the real time thermal propagators are having $2\times2$ matrix structure which corresponds to a doubling of degrees of freedom. This doubling of degrees of freedom in the real time  thermal field theory can be handled by considering the diagonal form of the propagators given by eqs.~\eqref{2.50} and \eqref{3.27}. In order to do so, we can start from the matrix form of the Dyson-Schwinger equation. The Dyson-Schwinger equation connects the interacting and the free thermal propagators through the self-energy matrix. Denoting the bosonic free propagator, interacting propagator and the self-energy matrices (denoted by bold-faced letters) 
as ${\bf D(k)}$, ${\bf D'(k)}$, and ${\bf \Pi}(k)$ respectively, the corresponding Dyson-Schwinger equation is given by,
\begin{equation}
\label{4.1}
{\bf D'}(k)={\bf D}(k) + {\bf D}(k){\bf \Pi}(k){\bf D'}(k).
\end{equation}
The matrix ${\bf U}(k_0)$, given by eq.~\eqref{2.51}, diagonalizes the free propagator ${\bf D}(k)$, the interacting propagator ${\bf D'}(k)$ and the self-energy ${\bf \Pi}(k)$, so that the  self-energy is of the form,
\begin{equation}
\label{4.2}
{\bf \Pi}(k)= {\bf U}(k_0)^{-1} \def\arraystretch{1 }
\begin{pmatrix}
\bar{\Pi}(k)  & 0\\
0 & - \bar{\Pi}(k)^*  \\
\end{pmatrix}{\bf U}(k_0)^{-1},
\end{equation} 
which, in turn, diagonalizes (denoted by 'bar') 
eq.~\eqref{4.1} and  reduces the matrix equation to an ordinary equation, 
\begin{equation}
\label{4.3}
\bar{D}=\triangle_F +\triangle_F\bar{\Pi }\bar{D},
\end{equation}    
having the solution 
\begin{equation}
\label{4.4}
\bar{D}=\frac{-1}{k^2-m^2+\bar{\Pi }},
\end{equation}
where $\bar{D}$ is the complete diagonalized thermal propagator and the $\bar{\Pi }$ is the diagonalized self-energy function.
It is evident from eq.~\eqref{4.2} that,
\begin{equation}
\label{4.5}
 \Pi_{22} =-\Pi^* _{11}, ~\Pi_{21} =\Pi _{12}.
\end{equation}
Also, the diagonalized self-energy $ \bar{\Pi } $ can be obtained entirely from any one component, say $\Pi _{11}$, from eq.~\eqref{4.2} as,
 \bea
 \label{4.6}
 \text{Re}\bar{\Pi } &=& \text{Re} \Pi_{11},~\text{and} \nn\\
 \text{Im}\bar{\Pi } &=& \epsilon(k_0)\tanh \left[\frac{q}{2(q-1)}\ln\left[1+\beta(q-1)k_0\right]\right]
 \text{Im}\Pi_{11}. \nn\\
 \label{relationdirac}
 \eea
 
 In a similar way, for the Dirac propagators, the Dyson-Schwinger equation reads,
 \begin{equation}
 \label{4.7}
 {\bf S'}(p)={\bf S}(p) + {\bf S}(p){\bf \Sigma}(p){\bf S'}(p).
 \end{equation}
The matrix ${\bf V}(p_0)$ given by eq.~\eqref{3.28}, which diagonalizes the free propagator 
${\bf S}(p)$, also diagonalizes the full propagator ${\bf S'}(p )$ and the self-energy ${\bf \Sigma}(p)$. Then the form of ${\bf \Sigma}(p)$ is,
 \begin{equation} 
 \label{4.8}
 {\bf \Sigma}(p)={\bf V}(p_0)^{-1}\def\arraystretch{1 }
 \begin{pmatrix}
 \bar{\Sigma}(p)  & 0\\
 0 & - \bar{\Sigma}(p)^*  \\
 \end{pmatrix}{\bf V}(p_0)^{-1},
 \end{equation} 
 which  reduces the matrix equation \eqref{4.7} to an ordinary equation,
 \begin{equation}
 \label{4.9}
 \bar{S}=S +S\bar{\Sigma }\bar{S},
 \end{equation}  
 with the solution 
 \begin{equation}
 \label{4.10}
 \bar{S}=\frac{-1}{\slashed{p}-m+\bar{\Sigma }}.
 \end{equation}
 It is evident from eq.~\eqref{4.8} that,
 \begin{equation}
 \label{4.11}
 \Sigma_{22}=-\Sigma^* _{11}, ~\Sigma_{21}=-\Sigma _{12}.
 \end{equation}
 \\
 The diagonalized self-energy $ \bar{\Sigma } $ can be obtained entirely from any one component, say $\Sigma _{11}$, from eq.~\eqref{4.8} as,
 \bea
 \label{4.12}
 \text{Re}\bar{\Sigma } &=& \text{Re} \Sigma_{11},~\text{and} \nn\\
 \text{Im}\bar{\Sigma } &=& \epsilon(p_0)\coth \left[\frac{q}{2(q-1)}\ln\left[1+\beta(q-1)p_0\right]\right]
 \text{Im}\Sigma_{11}. \nn\\
 \label{relationdirac}
 \eea
\section{Application: thermal mass of the scalar particles}
  \label{sec:tsallismD}
  
\begin{figure}[tbp]
\centering 
\hspace*{3cm} \includegraphics[width=.65\textwidth,trim=0 276 0 200,clip]{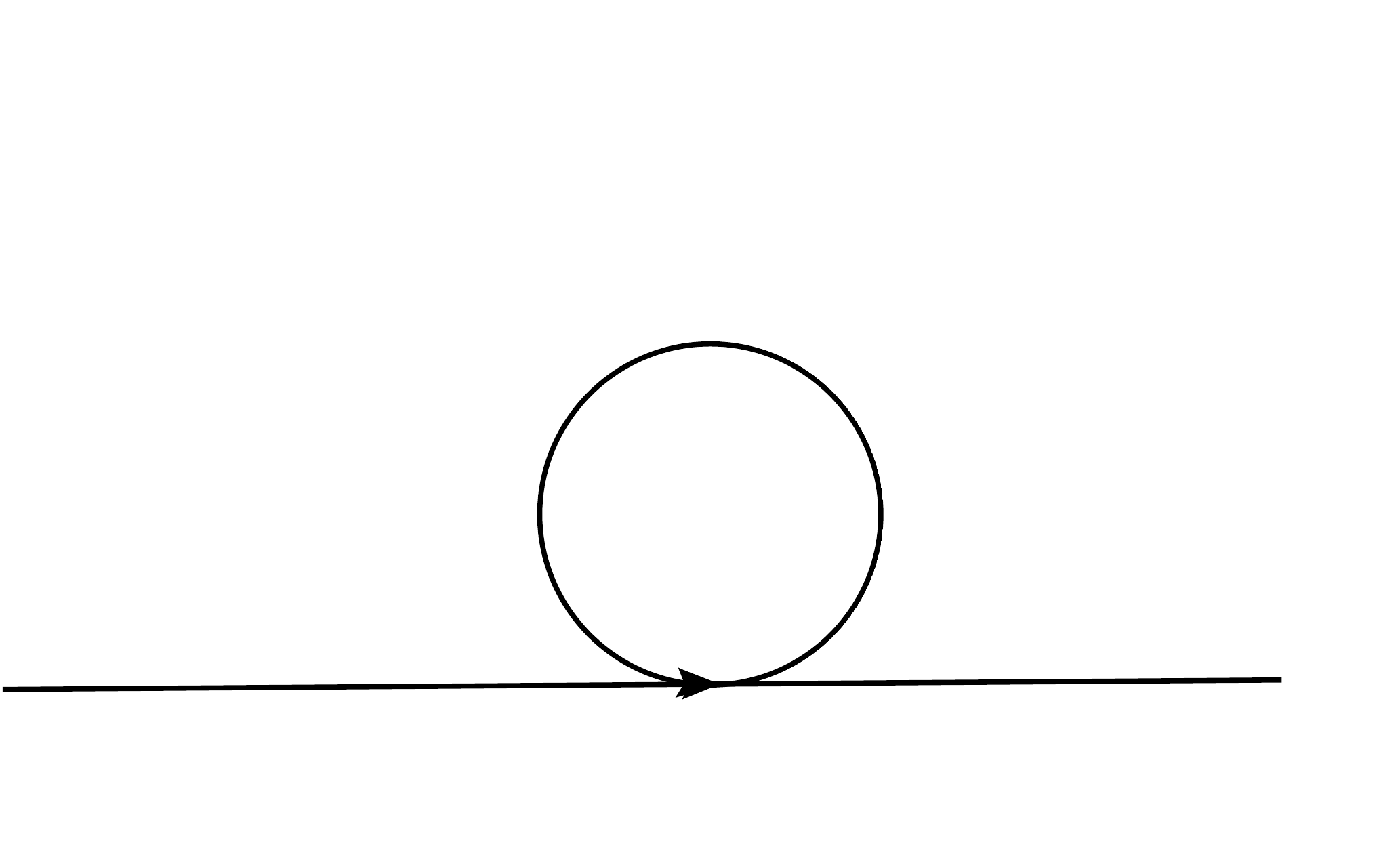}
\hfill
\caption{\label{fig2} Leading self-energy diagram in $\phi^4$ theory.}
\end{figure}

In this section we will compute the thermal mass of the scalar particles as an application of the
thermal field theory formalism developed so far for the Tsallis statistics. Thermal mass is given by the real
part of self-energy, which we calculate using the `11' component of the Tsallis thermal propagator.
We calculate the thermal mass in the self-interacting $\phi^4$ 
theory for which the Lagrangian density is given by,
 \begin{equation}
 \mathcal{L}=\frac{1}{2}\partial_\mu\phi\partial^\mu \phi-\frac{1}{2} m^2\phi^2-\frac{\lambda}{4 !}\phi^4.
 \end{equation}
The leading order mass shift due to thermal effects $\Delta m_{\text{T}}^2$, which involves the diagram shown in Fig.~\ref{fig2}, can be obtained as \cite{das},
 \begin{eqnarray}
  \triangle m_{\text{T}}^2 
  &=& \frac{\lambda}{2} \int\frac{d^3 p}{(2\pi)^3}\frac{1}{\omega_p} \frac{1}{\left[1+(q-1) \beta \omega_p\right]^{\frac{q}{q-1}}-1} \nonumber\\
  &=& \frac{ \lambda}{2} \sum_{s=1}^{\infty}\int\frac{d^3 p}{(2\pi)^3}\frac{1}{\omega_p} \left[1+(q-1) \beta \omega_p\right]^{-\frac{qs}{q-1}} \nonumber\\
  &=& \frac{ \lambda m^2}{4\pi^2} \sum_{s=1}^{\infty} \int dk \frac{k^2(1+k^2)^{-\frac{1}{2}}}{\Big\{1+(q-1) \frac{m}{T} \sqrt{1+k^2}\Big\}^{\frac{qs}{q-1}}}, \nn\\
  \label{thmass}
 \end{eqnarray}
 where $k=p/m$. The infinite sum converges for $\left[1+(q-1) \beta \omega_p\right]>1$. This condition is satisfied
 because in this work we take $q>1$ and consider massive particles.
We use the following Mellin-Barnes representation (for details see \cite{BhattaCleMog,smirnov}) in the last line of the above equation,
 \begin{eqnarray}
 \frac{1}{(X+Y)^{\lambda}} = \int_{\epsilon-i\infty}^{\epsilon+i\infty} \frac{dz}{2\pi i} \frac{\Gamma(-z)\Gamma(\lambda+z)}{\Gamma(\lambda)} \frac{Y^z}{X^{\lambda+z}}.
 \end{eqnarray}
 for $Y=1$, $X=(q-1)\sqrt{1+k^2}m/T$ and $\lambda=qs/(q-1)$, where $\mathrm{Re(\lambda)>0}$ and $\mathrm{Re}(\epsilon) \in (-\mathrm{Re}(\lambda),0)$, and obtain,
 \begin{eqnarray}
\triangle m_{\text{T}}^2 &=& \frac{ \lambda m^2}{4\pi^2} \sum_{s=1}^{\infty} \int_{\epsilon-i\infty}^{\epsilon+i\infty} \frac{dz}{2\pi i} 
\frac{\Gamma(-z)\Gamma\left(\frac{qs}{q-1}+z\right)}{\Gamma\left(\frac{qs}{q-1}\right) \Big\{(q-1)\frac{m}{T}\Big\}^{\frac{qs}{q-1}+z}} 
\int_{0}^{\infty} dk ~k^2 (1+k^2)^{-\frac{1}{2}-\frac{z}{2}-\frac{qs}{2(q-1)}} \nonumber\\
&=& \frac{ \lambda m^2}{16\pi^{\frac{3}{2}}} \sum_{s=1}^{\infty} \int_{\epsilon-i\infty}^{\epsilon+i\infty} \frac{dz}{2\pi i} 
\frac{\Gamma(-z)\Gamma\left(\frac{qs}{q-1}+z\right)}{\Gamma\left(\frac{qs}{q-1}\right) \Big\{(q-1)\frac{m}{T}\Big\}^{\frac{qs}{q-1}+z}} 
\frac{\Gamma\left(\frac{qs}{2(q-1)}+\frac{z}{2}-1\right)}{\Gamma\left(\frac{qs}{2(q-1)}+\frac{z}{2}+\frac{1}{2}\right)}.
 \end{eqnarray}
 The above expression has poles at the positive and negative $z$ values given by $z=\mathcal{K},-\lambda-\mathcal{K},~\mathrm{and~at}~z=2-2\mathcal{K}-\lambda~(\mathcal{K}\in \mathcal{Z}^{\geq})$ due to
 the appearance of the gamma functions in the numerator. We wrap the contour clockwise to include the poles at $z=\mathcal{K}$. While doing so, 
 the condition $(q-1)m/T>1$ guarantees the convergence of the integral at $z\rightarrow \infty$ . Summing up the residues due to the poles at $z=\mathcal{K}$, we obtain the following expression in terms of the summation of an infinite series whose terms contain the hypergeometric functions $_2F_1$ \cite{arfken},
 
 \begin{eqnarray}
\triangle m_{\text{T}}^2 &=& \frac{ \lambda m^2}{16\pi^{\frac{3}{2}}} \mathlarger{\mathlarger{\sum}}_{s=1}^{\infty}\left\{\frac{m (q-1)}{T}\right\}^{-\frac{q s}{q-1}} \left[\frac{\Gamma
   \left(\frac{q (s-2)+2}{2 (q-1)}\right) \, _2F_1\left(\frac{q
   (s-2)+2}{2 (q-1)},\frac{q s}{2 (q-1)};\frac{1}{2};\frac{T^2}{m^2
   (q-1)^2}\right)}{\Gamma \left(\frac{s q+q-1}{2
   (q-1)}\right)} \right.\nonumber\\
   &&-\left. \frac{T}{m(q-1)} \frac{\Gamma \left(\frac{q (s-1)+1}{2 (q-1)}\right)
   \Gamma \left(\frac{s q+q-1}{q-1}\right) \, _2F_1\left(\frac{q
   (s-1)+1}{2 (q-1)},\frac{s q+q-1}{2 (q-1)};\frac{3}{2};\frac{T^2}{m^2
   (q-1)^2}\right)}{\Gamma \left(\frac{q s}{q-1}\right) \Gamma
   \left(\frac{q (s+2)-2}{2 (q-1)}\right)}\right]. \nn\\
  \end{eqnarray}

\begin{figure}[t]
 	\centering	  
 	\includegraphics[width=8cm, height=7.0cm] {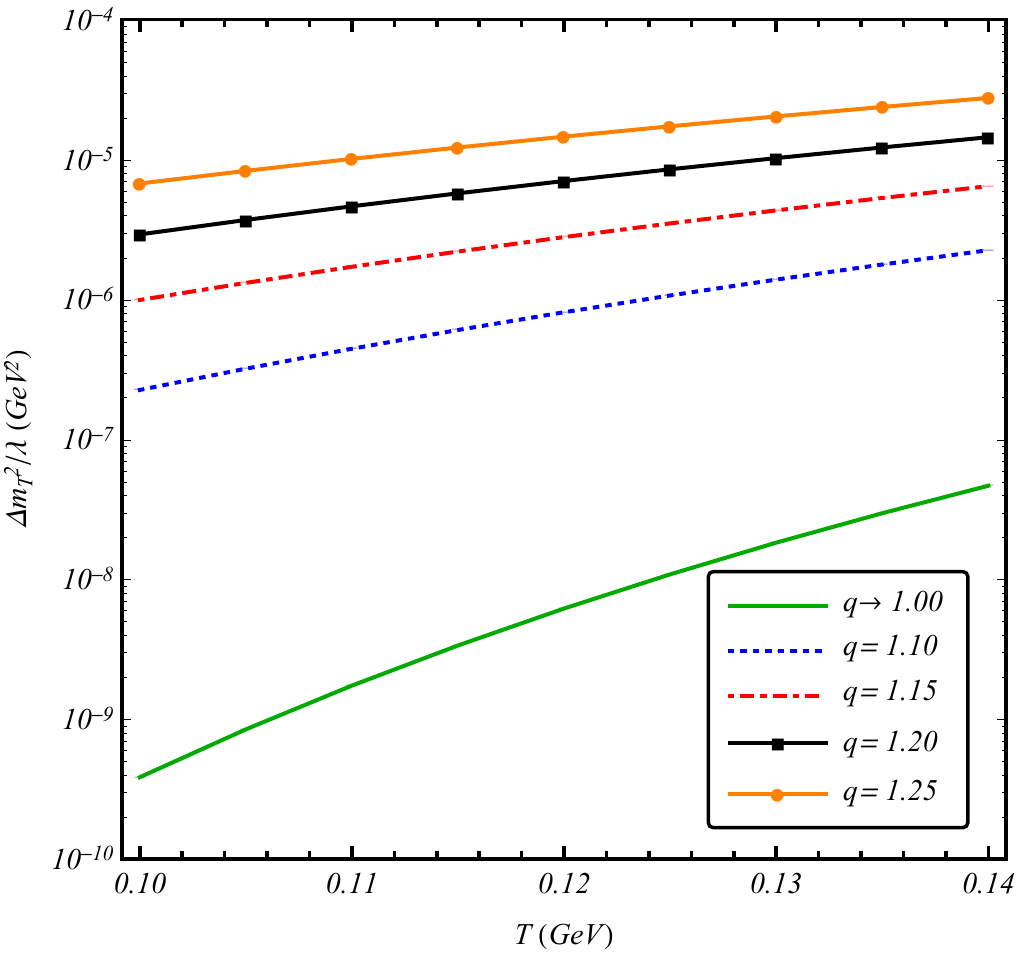}
 	\caption{Variation of the thermal mass of a scalar particle of mass 1.5 GeV with temperature for the Boltzmann-Gibbs (green,
	solid line) and the Tsallis statistics.}
 	\label{fig3}
 \end{figure}

Though the series appears to contain an infinite number of terms, we have verified that for the $q$, $T$, and mass values used
in Fig.~\ref{fig3}, the terms in the series die down very fast with increasing $s$. So, we may choose an 
upper cut-off $s=s_{\mathrm{max}}$ (convergence is obtained here for $s_{\text{max}}=3$ here) and
 obtain the following expression for the thermal mass in the $\phi^4$ theory within the scope of the Tsallis statistics,
  \begin{eqnarray}
  \triangle m_{\text{T}}^2 &=& \frac{ \lambda m^2}{16\pi^{\frac{3}{2}}} \mathlarger{\mathlarger{\sum}}_{s=1}^{s_{\mathrm{max}}}\left\{\frac{m (q-1)}{T}\right\}^{-\frac{q s}{q-1}} \left[\frac{\Gamma
   \left(\frac{q (s-2)+2}{2 (q-1)}\right) \, _2F_1\left(\frac{q
   (s-2)+2}{2 (q-1)},\frac{q s}{2 (q-1)};\frac{1}{2};\frac{T^2}{m^2
   (q-1)^2}\right)}{\Gamma \left(\frac{s q+q-1}{2
   (q-1)}\right)} \right.\nonumber\\
   &&-\left. \frac{T}{m(q-1)} \frac{\Gamma \left(\frac{q (s-1)+1}{2 (q-1)}\right)
   \Gamma \left(\frac{s q+q-1}{q-1}\right) \, _2F_1\left(\frac{q
   (s-1)+1}{2 (q-1)},\frac{s q+q-1}{2 (q-1)};\frac{3}{2};\frac{T^2}{m^2
   (q-1)^2}\right)}{\Gamma \left(\frac{q s}{q-1}\right) \Gamma
   \left(\frac{q (s+2)-2}{2 (q-1)}\right)}\right]. \nn\\
   \label{thermmassinf}
  \end{eqnarray}
 
 \noindent For $q\rightarrow1$, the above integral approaches the Boltzmann-Gibbs result \cite{das,kapusta}, 
 \begin{equation}
   \triangle m_{\text{T}}^2=  \frac{  \lambda T^2}{24}+\mathcal{O}(m/T).
 \end{equation}
\\
The variation of the quantity $\triangle m_{\text{T}}^2/\lambda$ given by eq.~\eqref{thermmassinf} with $T$ is depicted in Fig.\ref{fig3}. It is clear from the results 
displayed that in case of the Tsallis statistics, the magnitude of thermal mass is larger than that in the Boltzmann-Gibbs (BG) statistics. As expected, in the limit $q\rightarrow 1$, the two
results tend to merge with each other. It is important to mention at this point that in a small system
like quark gluon plasma (QGP), where applications of the BG statistics is uncertain, the thermal field theoretic 
formulation developed in this work for the Tsallis statistics will have crucial importance. The thermal spectral functions
of the quarks and gluons \cite{kyagi} estimated using the BG and Tsallis statistics 
will differ (as is the case for $\phi^4$ interaction) 
which may have significant impact on the signals of QGP~\cite{cywong}.    
 \\
 \\
\section{Summary and conclusions}
\label{sec:summaconclu}
In this work we have derived the real time thermal propagators for scalar and Dirac fields
within the scope of the Tsallis statistics and calculated the thermal mass of a scalar particle subjected 
to $\phi^4$ interaction using the generalized thermal two-point function. It has been observed that the quantum 
Tsallis distributions given by eqs.~\eqref{TFD}, and~\eqref{TBE} appear in the thermal
part of the Tsallis two-point functions. We also observe that when the entropic parameter $q$ approaches 1,
the classical and quantum Boltzmann-Gibbs thermal propagators are recovered. In the calculations of the thermal mass
we observe a significant increase in the scaled mass shift which approaches the Boltzmann-Gibbs limit as $q$ approaches 
1. Hence, the present work may be seen as a stepping stone towards the generalization of the Boltzmann-Gibbs finite temperature quantum field theory developed with the aid of the Tsallis statistical mechanical formulations. As indicated, this formalism and its extension may help one to compute the quantities like thermal mass, decay rate, energy loss in more realistic situations dealing with fluctuating ambience, and/or long-range correlations.

However, the calculation uses an approximation given by eq.~\eqref{factapprox} which may limit its applicability, 
and in turn, that of the distributions given by eqs.~\eqref{tsallisMBn0},~\eqref{TFD}, and~\eqref{TBE}.
As already mentioned in the introduction, these distributions are the approximate forms of the Tsallis single particle
distributions, and a more general formula for the propagators is required. In this article, we, however, were interested
in the approximate forms of the distributions, and their connection with a thermal quantum field theory. 
The connection between the most general form of the Tsallis single particle distribution \cite{Parvan19} and a thermal
quantum field theory will be a subject matter of our next work. In addition to that, the definition of the expectation values 
is according to the choice made in deriving the Tsallis distributions in eqs.~\eqref{tsallisMBn0},~\eqref{TFD}, and~\eqref{TBE}. A more consistent definition of the expectation values needs to be incorporated in future, too.

\acknowledgments
M.R. would like to thank the Department of Atomic Energy, Govt. of India for
financial support. T.B. thanks Bogoliubov Laboratory of Theoretical Physics, JINR for the travel grant to visit 
Variable Energy Cyclotron Centre (VECC), Kolkata, India where this work began. He also thanks VECC for 
the support extended to him during his visit. The authors thank Dr. Alexandru Parvan for very fruitful discussions. 



\end{document}